\documentclass{article}
\usepackage{pre_ltwol2e}
\usepackage{psfig}
\usepackage{epsfig}

\def\Journal#1#2#3#4{{#1} {\bf #2}, #3 (#4)}

\def\NPB{{\em Nucl. Phys.} B}

\def\PRL{\em Phys. Rev. Lett.}
\def\PRD{{\em Phys. Rev.} D}

\def\be{\begin{equation}}
\def\ee{\end{equation}}
\def\bea{\begin{eqnarray}}
\def\eea{\end{eqnarray}}

\begin{document}

\begin{flushright}
{\bf Fermilab-Pub-98/303-E} \\
\end{flushright}
\vskip 0.5cm
\begin{center} 
\Large \bf VECTOR BOSON PAIR PRODUCTION AND TRILINEAR GAUGE BOSON COUPLINGS - 
RESULTS FROM THE TEVATRON
\end{center}
\vskip 1.0cm
\centerline{H. T. Diehl}
\centerline{For the CDF + D\O \ Collaborations}
\centerline{\it Fermi National Accelerator Laboratory, Batavia, IL 60510}
\centerline{(September 25, 1998)}
\vskip 1.5cm
\centerline{\Large Abstract}
\vskip 1.0cm
Direct measurements of vector boson pair 
production processes and trilinear gauge boson couplings have been conducted 
by the CDF and D\O \ Collaborations.  
Preliminary results from searches for anomalous $WW/WZ\rightarrow \mu\nu$ 
jet jet and $WZ\rightarrow {\rm eee}\nu$ production are presented. 
95\% CL anomalous coupling limits 
from previously published D\O \ results are $-0.20 < \lambda < 0.20 \; 
(\Delta\kappa=0)$ and $-0.30 < \Delta\kappa < 0.43 \; (\lambda=0)$ for 
$\Lambda=2000$ GeV where the $WW\gamma$ couplings are assumed to equal the 
$WWZ$ couplings.  
Combined D\O \ + LEP experiment anomalous coupling limits are presented for
the first time. 95\% CL limits are $-0.16<\lambda_{\gamma}<0.10\; 
(\Delta\kappa=0)$ and $-0.15<\Delta\kappa_{\gamma}<0.41\; (\lambda=0)$
under the assumption that the couplings are related  by the ``HISZ'' 
constraints. 
95\% CL anomalous $ZZ\gamma$ and $Z\gamma\gamma$ coupling limits from D\O \
are $|h_{30}^{Z}|<0.36 \; (h_{40}^{Z}=0)$ and $|h_{40}^{Z}|<0.05 \;
(h_{30}^{Z}=0)$ for $\Lambda=750$ GeV. 
CDF reports the first observation of a $ZZ$ event. 
Prospects for Run~II are discussed.

\vskip 2.0in

\begin{flushleft}
------ \\
Submitted to the Proceedings of \\
ICHEP 98  XXIX International Conference on High Energy Physics \\
UBC, Vancouver, B.C. \\ Canada \\
July 23-29, 1998
\end{flushleft}
\global\advance\count0 -1
\clearpage

\twocolumn
\title{VECTOR BOSON PAIR PRODUCTION AND TRILINEAR GAUGE BOSON COUPLINGS - 
RESULTS FROM THE TEVATRON}

\author{H. T. DIEHL}

\address{M.S. 357,\\
Fermi National Accelerator Laboratory,\\
Batavia, IL 60510, USA \\
E-mail: DIEHL@FNAL.GOV}   


\twocolumn[\maketitle\abstracts{ Direct measurements of vector boson pair 
production processes and trilinear gauge boson couplings have been conducted 
by the CDF and D\O \ Collaborations.  
Preliminary results from searches for anomalous $WW/WZ\rightarrow \mu\nu$ 
jet jet and $WZ\rightarrow {\rm eee}\nu$ production are presented. 
95\% CL anomalous coupling limits 
from previously published D\O \ results are $-0.20 < \lambda < 0.20 \; 
(\Delta\kappa=0)$ and $-0.30 < \Delta\kappa < 0.43 \; (\lambda=0)$ for 
$\Lambda=2000$ GeV where the $WW\gamma$ couplings are assumed to equal the 
$WWZ$ couplings.  
Combined D\O \ + LEP experiment anomalous coupling limits are presented for
the first time. 95\% CL limits are $-0.16<\lambda_{\gamma}<0.10\; 
(\Delta\kappa=0)$ and $-0.15<\Delta\kappa_{\gamma}<0.41\; (\lambda=0)$
under the assumption that the couplings are related  by the ``HISZ'' 
constraints. 
95\% CL anomalous $ZZ\gamma$ and $Z\gamma\gamma$ coupling limits from D\O \
are $|h_{30}^{Z}|<0.36 \; (h_{40}^{Z}=0)$ and $|h_{40}^{Z}|<0.05 \;
(h_{30}^{Z}=0)$ for $\Lambda=750$ GeV. 
CDF reports the first observation of a $ZZ$ event. 
Prospects for Run~II are discussed.}]

\section{Introduction}

The Standard Model of electroweak interactions makes precise predictions
for the couplings between gauge bosons due to the non-abelian gauge symmetry of
$SU(2)_L\otimes U(1)_Y$. These self-interactions are described by the triple
gauge boson (trilinear) $WW\gamma$, $WWZ$, $Z\gamma\gamma$, and 
$ZZ\gamma$ couplings and the quartic couplings.  Deviations of the couplings
from the Standard Model (SM) values would indicate the presence of new
 physical phenomena.

The $WWV \; (V = \gamma ~{\rm or}~ Z)$ vertices are described by a
general effective Lagrangian~\cite{Wcoupling}
with two overall couplings, $g_{WW\gamma} = -e$ and
$g_{WWZ} = -e \cdot \cot \theta_{W}$, and
six dimensionless couplings
$g_{1}^{V}$, $\kappa_V$, and $\lambda_V$ $(V = \gamma$ or $Z)$,
after imposing {\it C}, {\it P} and {\it CP} invariance.
Electromagnetic gauge invariance requires that $g_{1}^{\gamma} = 1$, which
we assume throughout this paper.
The SM Lagrangian is obtained by setting $g_1^{\gamma} = g_1^Z = 1$,
$\kappa_{V} = 1 (\Delta\kappa_{V} \equiv \kappa_V - 1 = 0)$ and
$\lambda_V = 0$.

A different set of parameters, motivated by $SU(2)\times U(1)$ gauge
invariance, has been used by the LEP collaborations~\cite{alpha}.
This set consists of three independent couplings
$\alpha_{B\phi}$, $\alpha_{W\phi}$ and $\alpha_W$:
$\alpha_{B\phi}\equiv \Delta\kappa_{\gamma} - \Delta g_1^Z \cos^{2}\theta_{W}$,
$\alpha_{W\phi}\equiv\Delta g_1^Z \cos^{2}\theta_{W}$ and
$\alpha_{W}\equiv\lambda_{\gamma}$.
The remaining $WWZ$ coupling parameters $\lambda_Z$ and $\Delta\kappa_Z$
are determined by the relations $\lambda_Z = \lambda_{\gamma}$ and
$\Delta\kappa_Z = -\Delta\kappa_{\gamma}\tan^{2}\theta_{W} + \Delta g_1^Z$.
The HISZ relations~\cite{HISZ}
which have been used by the D{\O} and CDF collaborations are
also based on this set with the additional constraint
$\alpha_{B\phi} = \alpha_{W\phi}$.

The cross section with non-SM couplings grows with 
${\hat s}$.  In order to avoid unitarity violation, the 
anomalous couplings are modified as form factors with a scale $\Lambda$;
$\lambda_{V}(\hat{s}) = {\lambda_{V} \over ( 1 + \hat{s}/\Lambda^{2})^{2}}$ 
and $\Delta\kappa_{V}(\hat{s}) = 
{\Delta\kappa_{V} \over ( 1 + \hat{s}/\Lambda^{2})^{2}}$.

The $Z\gamma V \; (V = \gamma ~{\rm or}~ Z)$ vertices are described by
a general vertex function~\cite{Zcoupling}
with eight dimensionless couplings
$h_{i}^{V} (i=1,4 ~; V=\gamma ~{\rm or}~ Z)$.
In the SM, all of $h_{i}^{V}$'s are zero.
The form factors for these vertices,
similar to the $WWV$ vertices, are
$h_{i}^{V}(\hat{s}) = {h_{i0}^{V} \over ( 1 + \hat{s}/\Lambda^{2})^{n}}$,
where $n = 3$ for $i = 1,3$ and $n = 4$ for $i = 2,4$.

Vector boson pair production provides sensitive ground for {\em direct tests} 
of the trilinear couplings.  This paper provides a description of recent 
results from D\O \ and CDF, including preliminary results from searches for 
anomalous couplings in $WW/WZ\rightarrow \mu\nu$ jet jet and
$WZ\rightarrow eee \nu$ final states at D\O, a description of a recently 
published, combined analysis, anomalous $WW\gamma$ and $WWZ$ coupling 
result from D\O,  
combined D\O \ + LEP anomalous $WW\gamma$ and $WWZ$ couplings results
(presented for the first time), limits on anomalous $Z\gamma\gamma$ and 
$ZZ\gamma$ couplings from studies of $Z\gamma$ final states at D\O, and 
report of the observation of a $ZZ\rightarrow \mu\mu\mu\mu$ event at CDF 
in Run~I. Prospects for Run~II are discussed.

\section{Preliminary Results from D\O}
\subsection{Search for anomalous $WW/WZ\rightarrow \mu\nu \; {\rm jet\; jet}$ 
Production}
This analysis searches for anomalous $WW$ or $WZ$ production using the decay
signature $W\rightarrow\mu\nu$ and $W/Z\rightarrow {\rm jj}$.  We cannot 
distinguish between $W$ and $Z$ decays to jets.  Further, because SM
$WW$ and $WZ$ production is swamped by backgrounds having this same signature,
the analysis is sensitive only to production with anomalous couplings. 

We selected events with an isolated, central muon with $p_T>20$ GeV/c and 
with \hbox{$\rlap{\kern0.25em/}E_T$}$>20$ GeV where the transverse mass 
$M_T(\mu\nu)>40$ GeV/c$^2$, indicative of a $W$ boson decay. We required that
the event contain at least two jets with $E_T>20$ GeV. For 224 of the 372 
candidates which survived those selection criteria, the invariant mass of the 
two highest $E_T$ jets was between $50$ and $110$ GeV/c$^2$, as expected from
the hadronic decay of a $W$ or $Z$ boson. From Monte Carlo simulation, we 
expected $4.04^{+0.54}_{-0.68}$ $WW$ events and $0.49^{+0.10}_{-0.11}$ $WZ$ 
events, given SM trilinear couplings.  Of course, if the couplings are non-SM,
there will be an excess of events with high-$p_T$ $W$ bosons.

\begin{figure}
\vspace{-.33in}
\psfig{figure=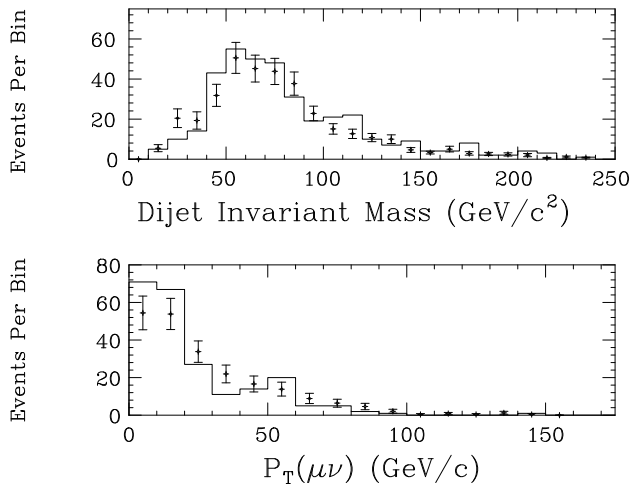,width=3.in}
\vspace{-1.in}
\caption{D\O \ 
$WW/WZ\rightarrow \mu\nu \; {\rm jet\; jet}$ analysis.  The upper plot is
the dijet invariant mass.  The lower plot is $p_T(\mu\nu)$ for events which 
pass the dijet mass selection criteria. The histogram is the data.  The points
are the background and it's uncertainty.}
\label{fig:wwmass}
\end{figure}

The background consists principally of $W+\ge 2$ jets ($117\pm24$ events 
expected) and multijet events with an accidentally isolated muon from a
heavy quark decay ($105\pm19$ events expected).  Small contributions to the 
background arise from $t\bar{t}$ production. The total expected background
is $224\pm31$ events.    Figure~\ref{fig:wwmass} shows the dijet invariant 
mass and $p_T(\mu\nu)$ for the data and expected background.  

Since we observed no excess of events in the high-$p_T(\mu\nu)$ region,
we place upper limits on anomalous $WW\gamma$ and $WWZ$ couplings.  This is
performed using a binned likelyhood fit of the observed $p_T(\mu\nu)$ spectrum
to the expected $p_T(\mu\nu)$ spectrum, given anomalous couplings plus
the expected background.  The 1-D 95\% CL coupling limits are
$-0.45<\lambda<0.46 \; (\Delta\kappa=0)$ and $-0.62<\Delta\kappa<0.78 \;
(\lambda=0)$ for $\Lambda=1500$ GeV, assuming the $WW\gamma$ couplings equal
the $WWZ$ couplings.  While previous analyses have produced more 
restrictive anomalous coupling limits, these results will ultimately 
contribute to the  D\O \ combined results.

\subsection{Search for $WZ\rightarrow eee\nu$ Events} 
Searches for anomalous $WZ$ production enable the possibility of constraining 
the $WWZ$ couplings independant of the $WW\gamma$ couplings.  D\O \ has 
searched for the process $WZ\rightarrow eee\nu$.  Two electrons with 
$E_T>25$ GeV, one more electron with $E_T>10$ GeV, and 
\hbox{$\rlap{\kern0.25em/}E_T$} $>15$ GeV are required in candidate events.
The transverse mass of one electron and the  \hbox{$\rlap{\kern0.25em/}E_T$}
must be greater the 30 GeV/c$^2$.  The invariant mass of the other two 
electrons must be between 81 and 101 GeV/c$^2$.  One candidate survives 
the selection criteria.  

\begin{figure}
\begin{center}
\psfig{figure=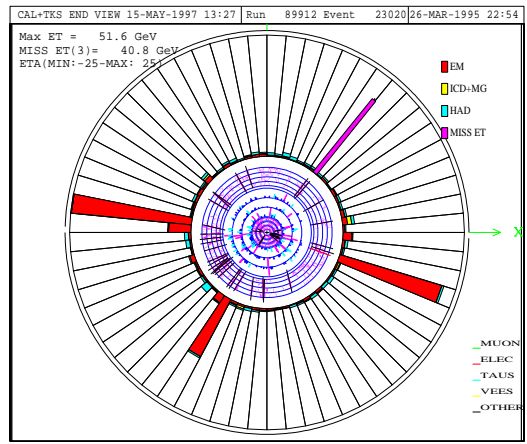,width=2.7in}
\end{center}
\caption{$WZ\rightarrow eee\nu$ candidate from D\O. The figure shows the 
$E_T$, represented by the height of the column, deposited in the central
calorimeter.  The three largest towers are the three electrons.  The narrow
tower shows the direction and magnitude of the missing transverse energy.}
\label{fig:wz_event}
\end{figure}

Based on SM Monte Carlo, $0.146\pm 0.011$ $WZ$ events are expected.  
The background, principally due to $Z+{\rm jet}$ events where the jet mimicks
an electron and there is accidental \hbox{$\rlap{\kern0.25em/}E_T$}, amounts
to $0.38\pm0.14$ events. Figure~\ref{fig:wz_event} shows the candidate's 
event display.  One pair of electrons has invariant mass of 93.6 GeV/c$^2$;
the transverse mass of the other electron and the missing $E_T$ is 
74.7 GeV/c$^2$. 

$WZ$ production is more sensitive to the value of $\lambda_Z$ and 
$\Delta g_1^Z$ than $\Delta\kappa_Z$.  Given one observed event and taking 
into account the expected background, 95\% CL limits are $|\lambda|<2.07
\;(\Delta g_1^Z=0)$ and $|\Delta g_1^Z|<2.56 \; (\lambda_Z=0)$ for 
$\Lambda=1000$ GeV.  Although the limits are looser than some of 
those previously
measured, they are independant of any assumptions on the relation between
$WW\gamma$ and $WWZ$ couplings. Ultimately, these results will help
constrain $\Delta g_1^Z$ in D\O's combined limits. 

\section{Combined $WW\gamma$ and $WWZ$ Anomalous Coupling Limits from D\O}
\begin{figure}
\hbox{
\epsfig{figure=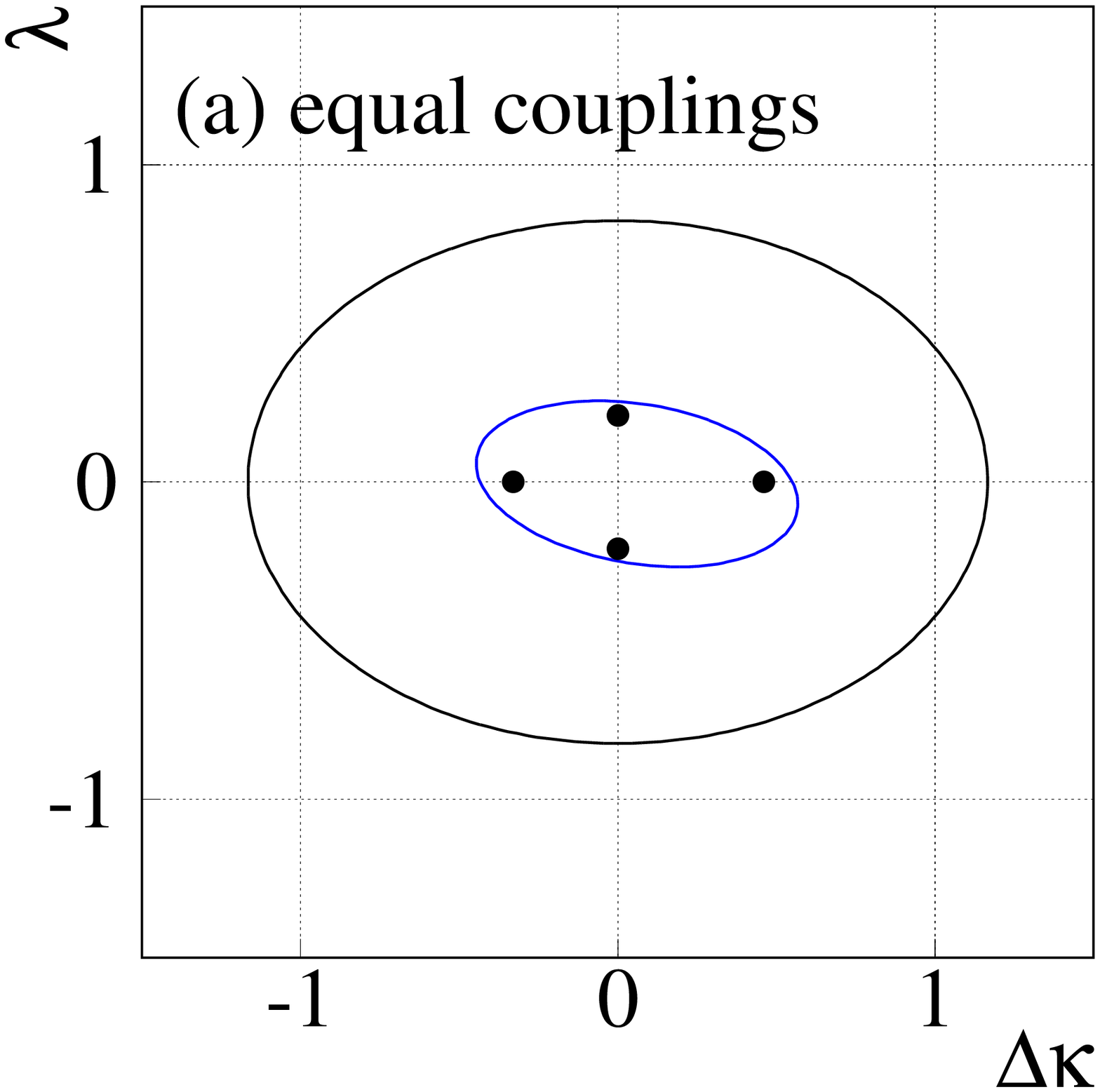,width=1.7in}
\epsfig{figure=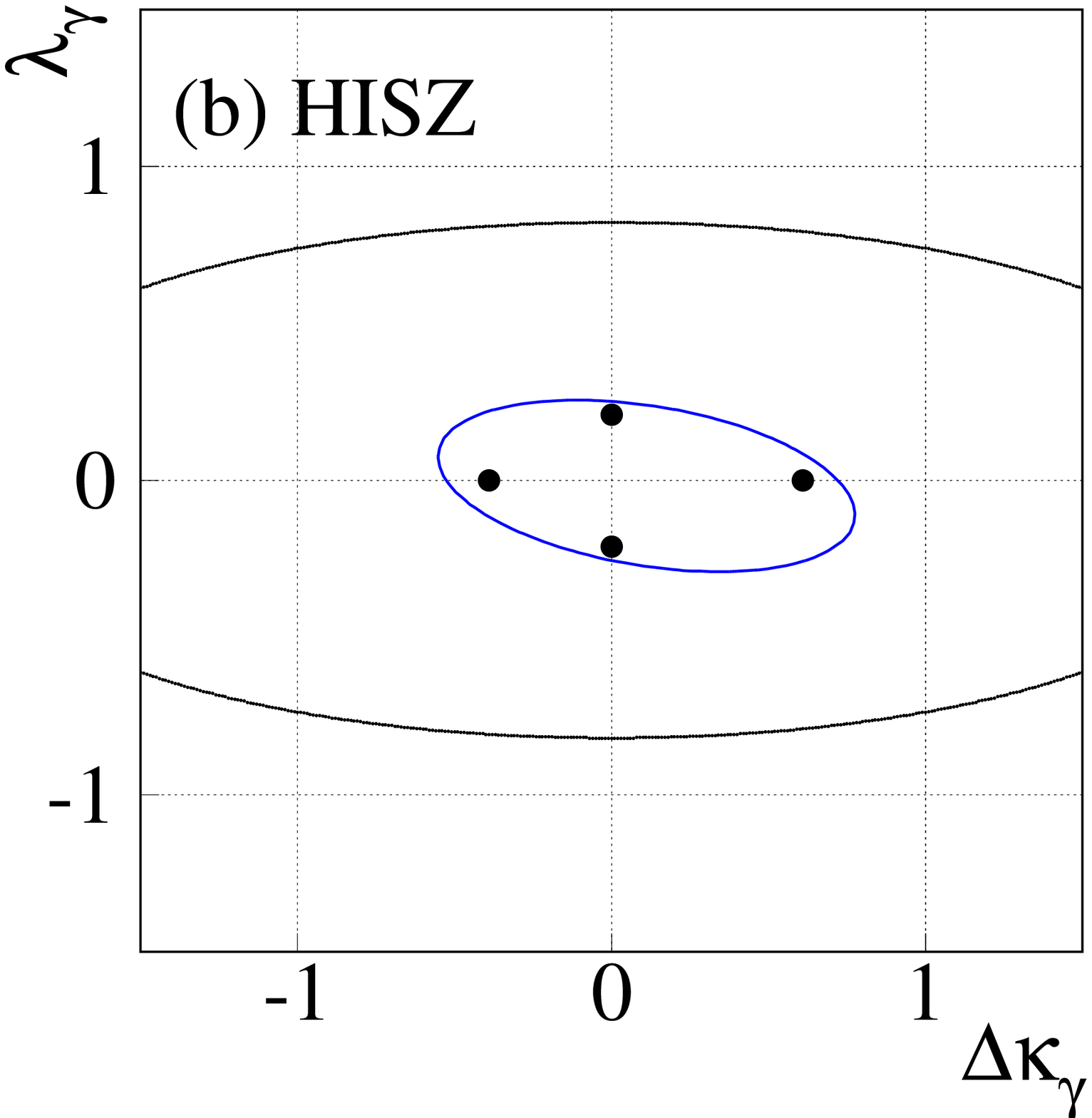,width=1.7in}}
\hbox{
\epsfig{figure=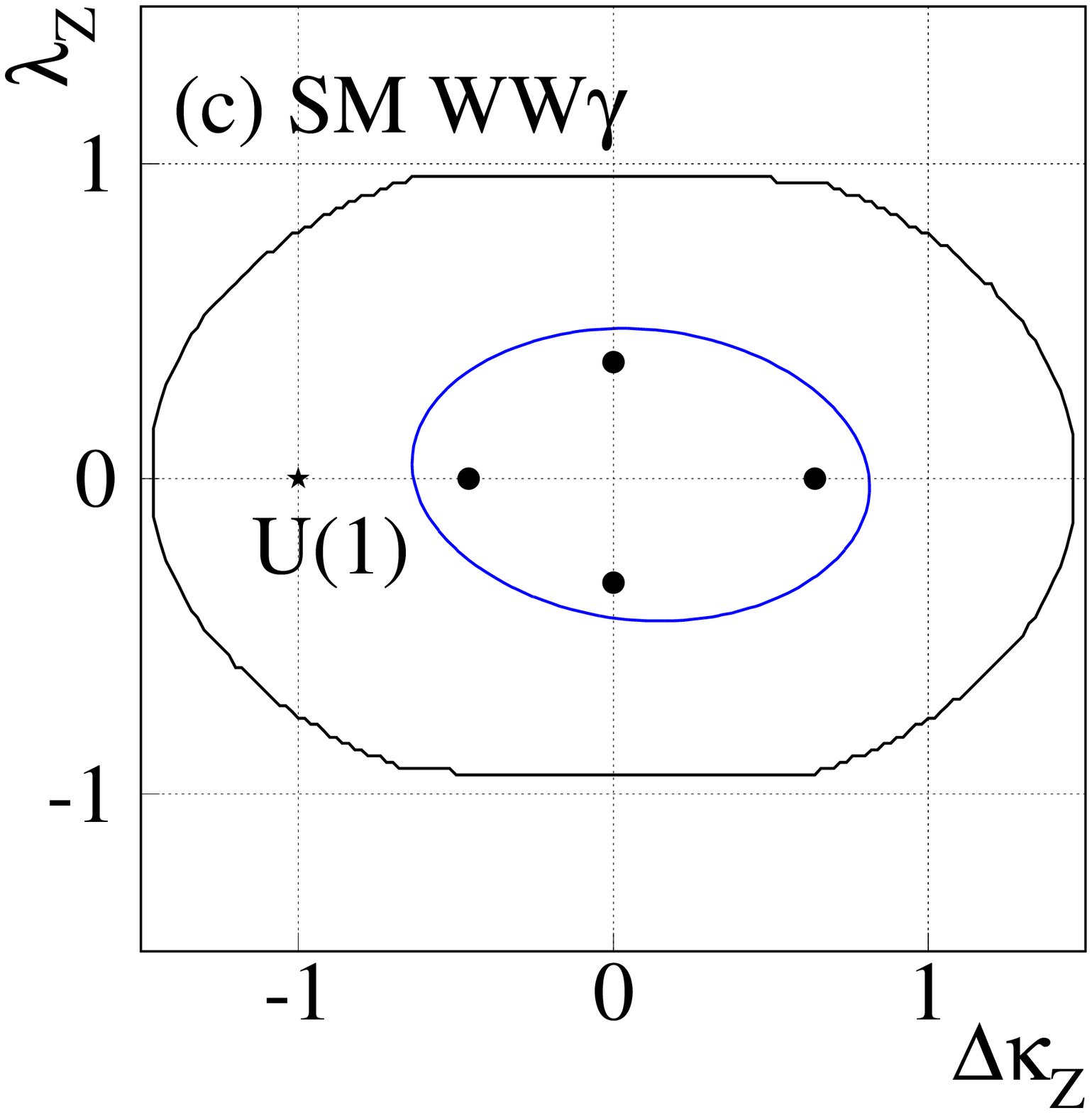,width=1.7in}
\epsfig{figure=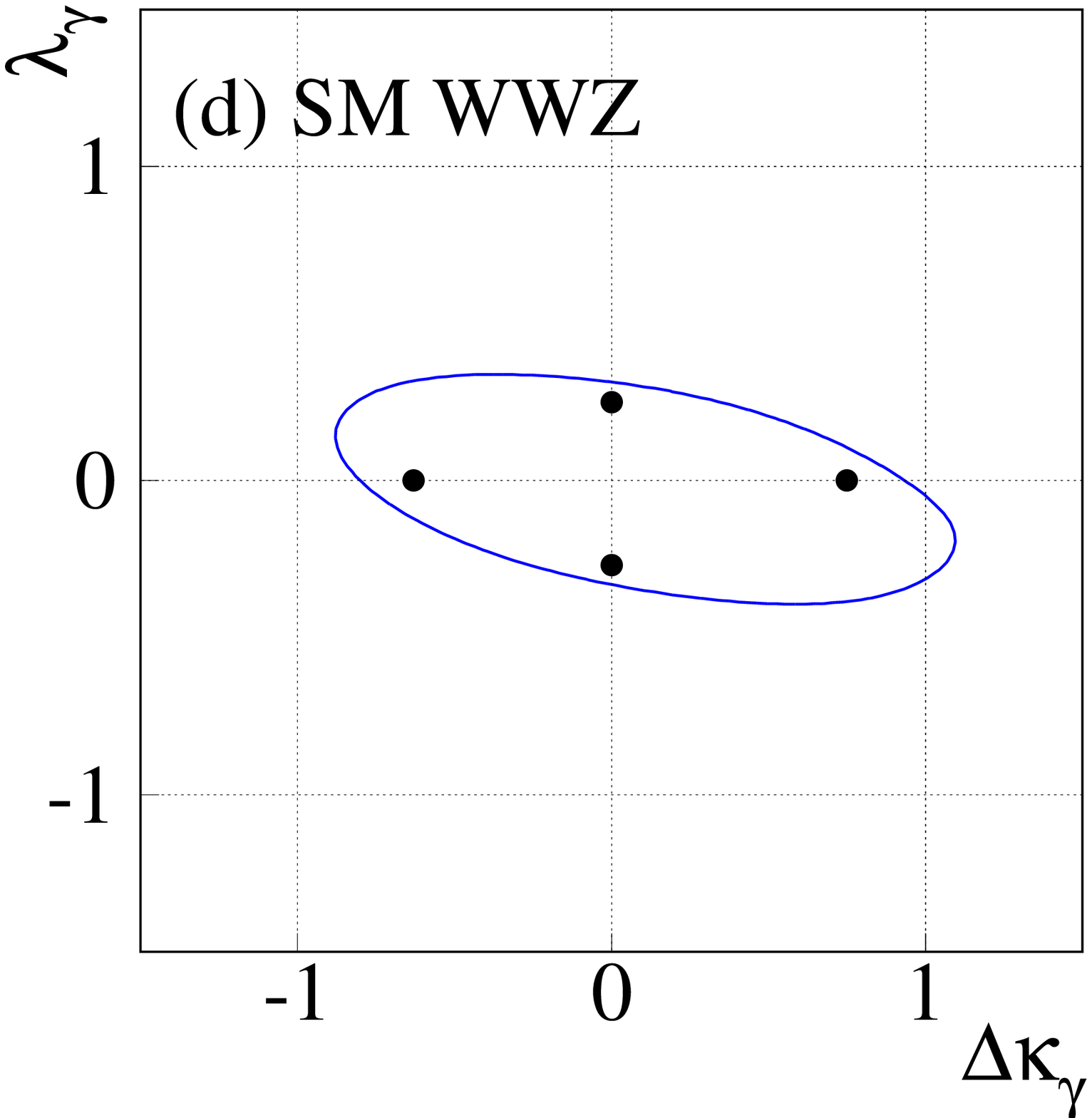,width=1.7in}}
 \caption{Contour limits on anomalous couplings
 from a simultaneous fit to the D\O \ data sets of $W\gamma$, $WW\rightarrow$
 dilepton, and $WW/WZ\rightarrow e\nu jj$ final states
 for $\Lambda = 1.5$ TeV:
 (a) $\Delta\kappa \equiv \Delta\kappa_\gamma = \Delta\kappa_Z,
 \lambda \equiv \lambda_\gamma = \lambda_Z$; (b) HISZ relations;
 (c) SM $WW\gamma$ couplings; and (d) SM $WWZ$ couplings.
 (a), (c), and (d) assume that $\Delta g_1^Z=0$.
 The solid circles correspond to $95 \%$ C.L. one-degree of freedom exclusion
 limits. 
 The inner and outer curves are
 $95 \%$ C.L. two-degree of freedom exclusion contour
 and the constraint from the unitarity condition, respectively.
 In (d), the unitarity contour is located outside of the
 boundary of the plot. The HISZ results include the 
 additional constraint $\alpha_{B\phi} = \alpha_{W\phi}$.}
\label{fig:combined}
\end{figure}

\begin{figure}[htb]
\centerline{\hbox{
\epsfig{figure=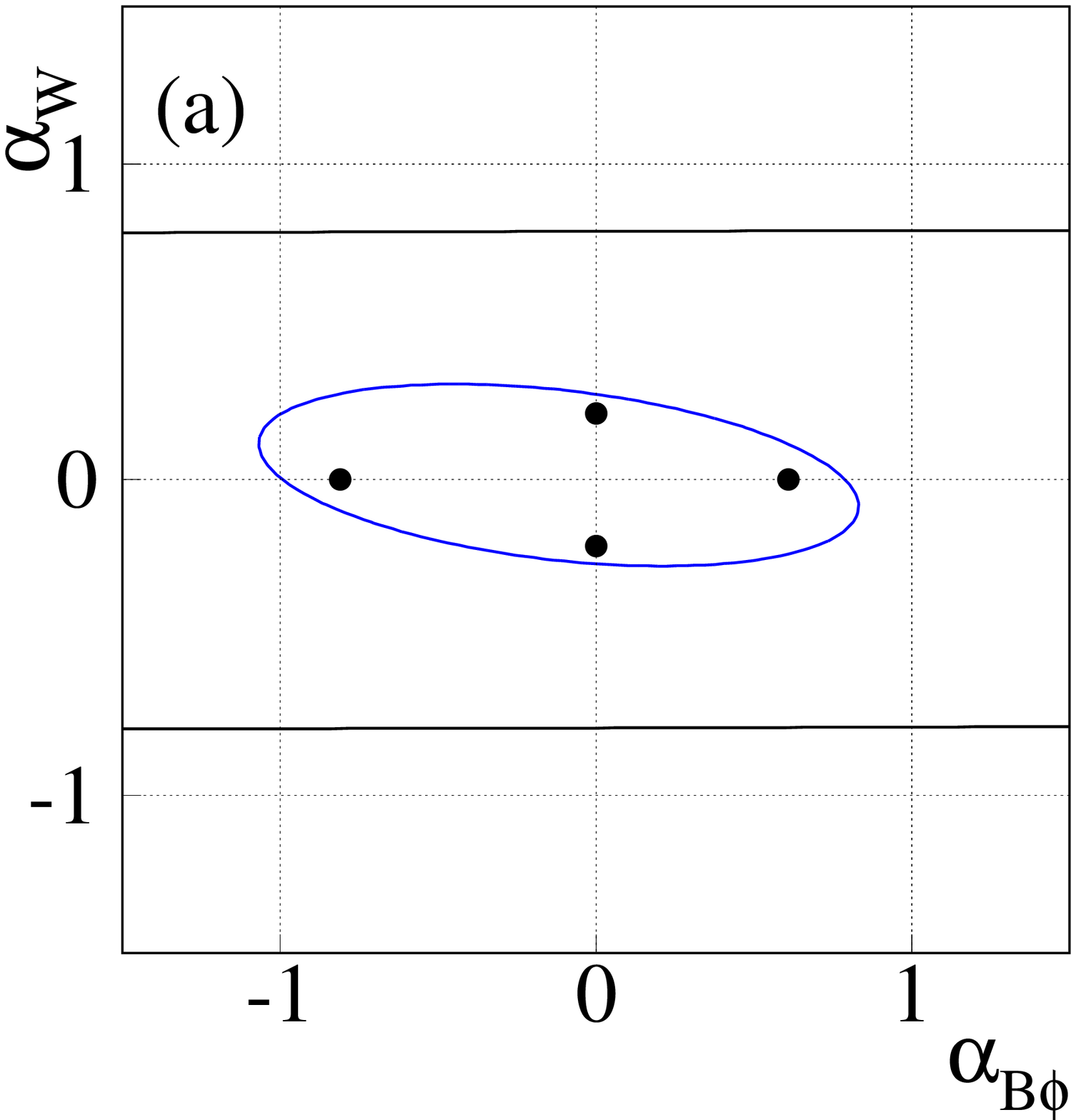,width=1.8in}
\epsfig{figure=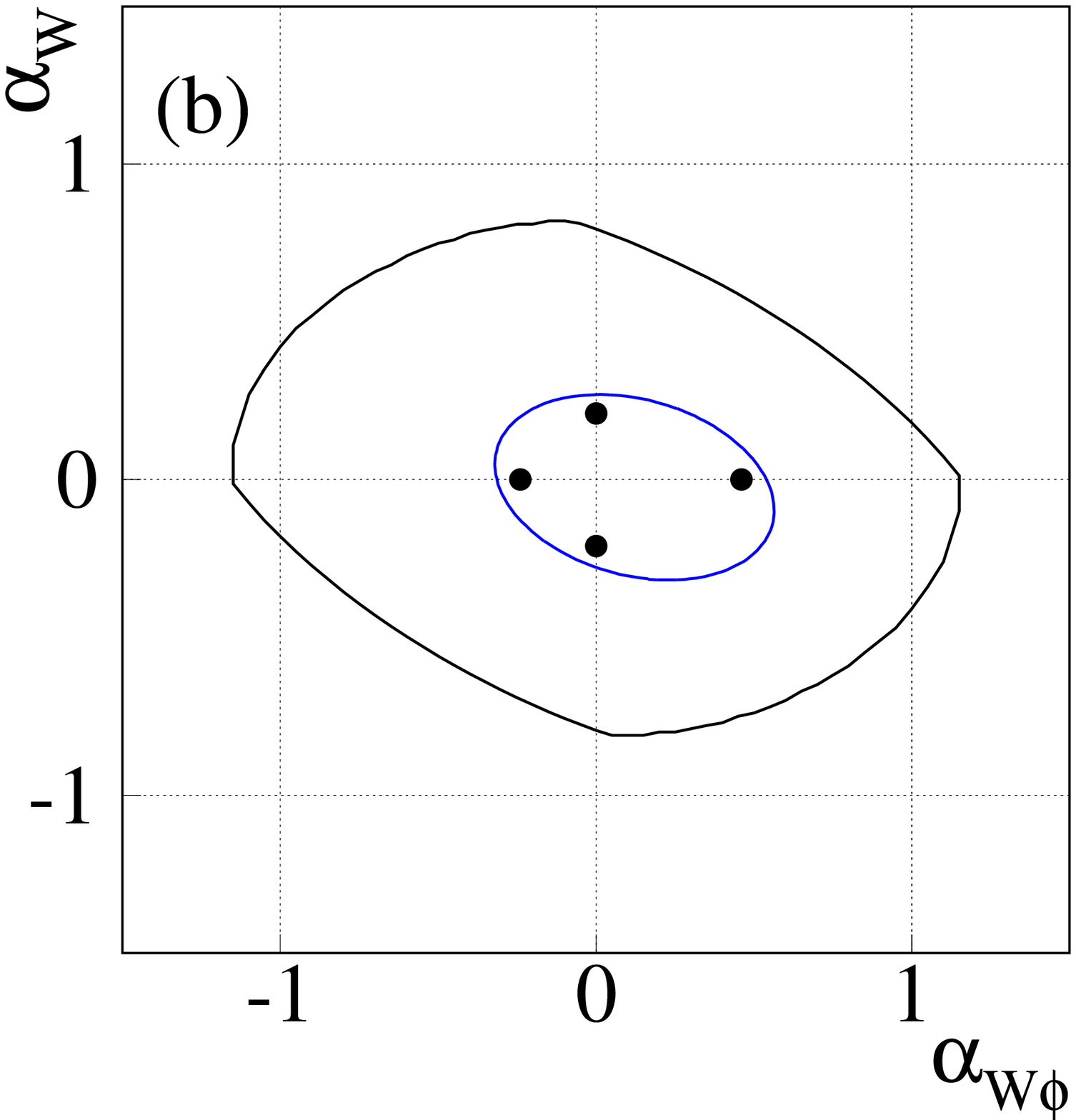,width=1.8in}}
}
 \caption{Contour limits on anomalous couplings
 from a simultaneous fit to the D\O \ data sets of the 
 $W\gamma$, $WW\rightarrow$
 dilepton, and $WW/WZ\rightarrow e\nu jj$ final states
 for $\Lambda = 1.5$ TeV:
 (a) $\alpha_{W}$ vs $\alpha_{B\phi}$ when $\alpha_{W\phi} = 0$; and
 (b) $\alpha_{W}$ vs $\alpha_{W\phi}$ when $\alpha_{B\phi} = 0$.
 The solid circles correspond to $95 \%$ C.L. one-degree of freedom exclusion
 limits. 
 The inner and outer curves are
 $95 \%$ C.L. two-degree of freedom exclusion contour
 and the constraint from the unitarity condition, respectively.}
 \label{fig:lep_sets}
\end{figure}

We have combined previously published limits on $WW\gamma$
couplings obtained from a fit to the photon $E_T$ spectrum in 
$W\gamma$ events~\cite{Wg1b_d0}, limits on $WWZ$ and $WW\gamma$ couplings
obtained from a fit to the $E_T$ of the two charged leptons in 
$WW\rightarrow{\rm dilepton}$ events~\cite{WW1b_d0}, and 
limits on $WWZ$ and $WW\gamma$ couplings obtained from a 
fit to the $p_T$ spectrum of the electron-neutrino system in 
$WW/WZ\rightarrow e\nu jj$ events~\cite{WWWZ1b_d0}. 
The results~\cite{d0_combined} are shown in Tables~\ref{table:combined_limits}
and \ref{table:combined_lims2} and in Figures~\ref{fig:combined} and 
\ref{fig:lep_sets}.
\begin{table*}[htb]
\begin{center}
\begin{tabular}{|c|c|c|} \hline
Coupling &$\Lambda=1.5$ TeV&$\Lambda=2.0$ TeV\\ \hline
$\lambda_{\gamma} = \lambda_Z$ ($\Delta\kappa_{\gamma}=\Delta\kappa_Z=0$)&
 $-0.21,~0.21$& $-0.20,~0.20$\\
$\Delta\kappa_{\gamma} = \Delta\kappa_Z$ ($\lambda_{\gamma}=\lambda_Z=0$)&
 $-0.33,~0.46$& $-0.30,~0.43$\\ \hline
$\lambda_{\gamma}$(HISZ) ($\Delta\kappa_{\gamma}=0$)&
 $-0.21,~0.21$& $-0.20,~0.20$\\
$\Delta\kappa_{\gamma}$(HISZ) ($\lambda_{\gamma}=0$)&
 $-0.39,~0.61$& $-0.37,~0.56$\\ \hline
$\lambda_Z$(SM $WW\gamma$) ($\Delta\kappa_Z=\Delta g^Z_1=0$)&
 $-0.33,~0.37$& $-0.31,~0.34$\\
$\Delta\kappa_Z$(SM $WW\gamma$) ($\lambda_Z=\Delta g^Z_1=0$)&
 $-0.46,~0.64$& $-0.42,~0.59$\\
$\Delta g^Z_1$(SM $WW\gamma$) ($\lambda_Z=\Delta\kappa_Z=0$)&
 $-0.56,~0.86$& $-0.52,~0.78$\\ \hline
$\lambda_{\gamma}$(SM $WWZ$) ($\Delta\kappa_{\gamma}=0$)&
 $-0.27,~0.25$& $-0.26,~0.24$\\
$\Delta\kappa_{\gamma}$(SM $WWZ$) ($\lambda_{\gamma}=0$)&
 $-0.63,~0.75$& $-0.59,~0.72$\\ \hline
\end{tabular}
\end{center}
\caption{Limits at 95\% C.L. from a simultaneous fit to the D\O \ $W\gamma$,
$WW\rightarrow$ dilepton and $WW/WZ\rightarrow e\nu jj$ data samples.
The four sets of limits apply the same assumptions as the four components
(a), (b), (c) and (d), respectively, of Fig.~\ref{fig:combined}.
The HISZ results include the 
 additional constraint $\alpha_{B\phi} = \alpha_{W\phi}$.}
\label{table:combined_limits}
\end{table*}

\begin{table}[htb]
\begin{tabular}{|c|c|c|} \hline
Coupling &$\Lambda=1.5$ TeV&$\Lambda=2.0$ TeV \\ \hline
$\alpha_{B\phi}$ ($\alpha_{W\phi}=\alpha_W=0$)&
 $-0.81,~0.61$& $-0.77,~0.58$ \\ \hline
$\alpha_{W\phi}$ ($\alpha_{B\phi}=\alpha_W=0$)&
 $-0.24,~0.46$& $-0.22,~0.44$ \\ \hline
$\alpha_W$ ($\alpha_{B\phi}=\alpha_{W\phi}=0$)&
 $-0.21,~0.21$& $-0.20,~0.20$ \\ \hline
$\Delta g^Z_1$ ($\alpha_{B\phi}=\alpha_W=0$)&
 $-0.31,~0.60$& $-0.29,~0.57$ \\ \hline
\end{tabular}
\caption{Limits at 95\% C.L. on $\alpha$ parameters from a simultaneous fit
to the D\O \ $W\gamma$, $WW\rightarrow$
dilepton and $WW/WZ\rightarrow e\nu jj$ data samples.}
\label{table:combined_lims2}
\end{table}

\vspace{-0.5in}

\section{Combined D\O \ and LEP Anomalous $WW\gamma$ and $WWZ$ Couplings}
\begin{figure}
\begin{center}
\epsfig{figure=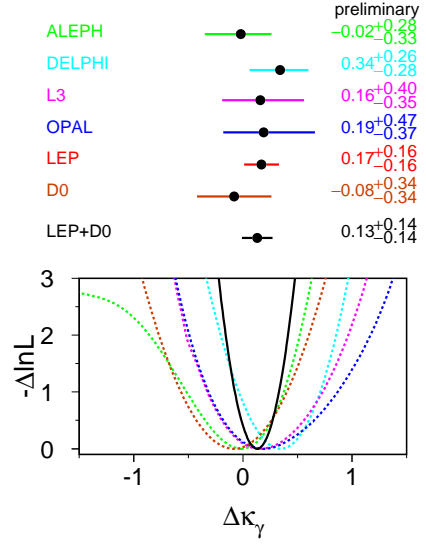,height=2.9in}
\end{center}
\caption{Limits on  $\Delta \kappa_{\gamma}$
and the log-likelyhood curves for the LEP experiments, D\O, and 
the combination, assuming 
$\lambda_{\gamma}=\lambda_Z=\Delta g_1^Z=0$ and that $\Delta\kappa_Z$ is 
determined from the HISZ relations.  At 95\% CL, the combined limit is 
$-0.15<\Delta\kappa_{\gamma}<0.41$. }
\label{fig:lep_set_1}
\end{figure}

\begin{figure}
\begin{center}
\epsfig{figure=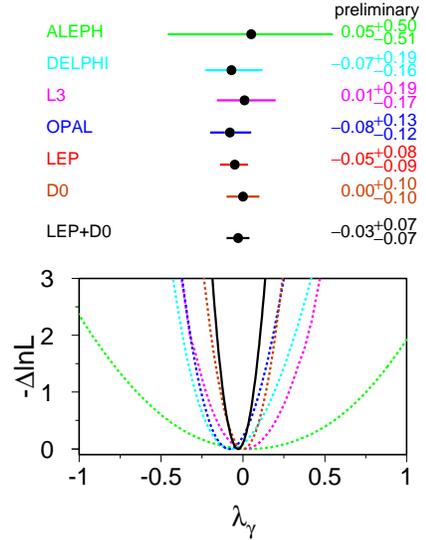,height=2.9in}
\end{center}
\caption{Limits on $\lambda_{\gamma}$
and log-likelyhood curves for the LEP experiments, D\O, and 
the combination, assuming  $\lambda_{\gamma}=\lambda_Z$,
and $\Delta\kappa_{\gamma}=\Delta\kappa_Z=0$. The limits are $68\%$ CL.
At 95\% CL, the combined limit is $-0.16<\lambda_{\gamma}<0.10$. }
\label{fig:lep_set_3}
\end{figure}

The published results from D\O \ and the four LEP experiments (see the 
summary~\cite{hywel} by Hywel Phillips) were combined to produce the tightest
available $WW\gamma$ and $WWZ$ coupling limits.  The procedure, documented
in detail in preprint form~\cite{lepd0}, is to add the negative 
log-likelyhoods presented as 
a function of trilinear coupling from the several LEP analyses and the 
combined D\O \ analysis described above.  The one standard
deviation limits are obtained directly from the curves by taking the values 
of the coupling where $\Delta \log{\cal{L}}=+0.5$ from the minimum. The 95\%
CL limit is given by the values of the coupling where 
$\Delta \log{\cal{L}}=+1.92$.  The dependencies on correlated systematic
errors and on $\Lambda$ are negligible. 
Figures~\ref{fig:lep_set_1} and \ref{fig:lep_set_3} show log-likelihoods 
and one $\sigma$ limits on $\lambda_{\gamma}$ and $\Delta\kappa_{\gamma}$ 
where one coupling is varied at a time and the $WWZ$ couplings are 
related to the $WW\gamma$ couplings through the HISZ equations
(without the extra constraint $\alpha_{B\phi} = \alpha_{W\phi}$).

The 95\% CL limits are $-0.16<\lambda_{\gamma}<0.10 \; (\Delta\kappa =0)$ 
and $-0.15<\Delta\kappa_{\gamma}<0.41 \; (\lambda=0)$. These limits provide
an update to the original analysis~\cite{lepd0} due to the inclusion 
of new LEP results. 


\section{$Z\gamma$ and $ZZ$ Production and Anomalous $ZZ\gamma$ and $Z\gamma
\gamma$ Couplings from D\O \ and CDF}
D\O \ has recently published~\cite{zgnng,zgeemmg} studies of $Z\gamma$ 
production using the $\nu\nu\gamma$, $ee\gamma$, and $\mu\mu\gamma$ final 
states.  The two neutrino final state makes up for it's smaller integrated 
luminosity by having a larger branching fraction than the charged lepton
mode and has the advantage that there are no photons radiating from the $Z$'s
decay products.  On the other hand, in the charged-lepton decay mode the 
backgrounds are smaller and the final state can be completely reconstructed.

The $Z\gamma\rightarrow \nu\nu\gamma$ analysis, with a minimum photon $E_T$ of
$40$ GeV, yielded 4 candidates with an expected background of $5.8\pm 1.0$ 
events in an integrated luminosity of $14$ pb$^{-1}$.  The analysis of the 
charged-lepton decay mode yielded 29 candidates with an expected background
of $5.4\pm 1.0$ events in $97$ pb$^{-1}$. It is interesting that there were 
two events in the charged-lepton sample with an $\sim 75$ GeV photon and with 
an $\ell^+ \ell^- \gamma$ invariant mass of $\sim 200$ GeV/c$^2$. 
SM $Z\gamma$ production would yield two or more events with $E_T^{\gamma}
>60 \;(70)$ GeV in 15\% (7.3\%) of repeats of the experiment. Also,
SM Monte Carlo indicates the most likely $Z\gamma$ mass for events with 
$E_T^{\gamma}$ in the range 70 to 79 GeV is $200$ GeV/c$^2$. CDF 
noted~\cite{cdfzg1a} a $Z\gamma$ event with $E_T^{\gamma}\sim 64$ GeV and 
$M(\mu\mu\gamma)\sim 188$ GeV/c$^2$ in their 20 pb$^{-1}$ sample from Run~Ia.
While there is no evidence of non-SM physics here, it is something to keep
one's eye on in the future.

The tightest limits on anomalous $ZZ\gamma$ and $Z\gamma\gamma$ couplings 
are those from the combined~\cite{zgeemmg} D\O \ analyses.  At 95\% CL,
the $Z\gamma\gamma$ limits are $|h_{30}^{\gamma}\; (h_{10}^{\gamma})|<0.37$ 
and $|h_{40}^{\gamma}\; (h_{20}^{\gamma})|<0.05$ with 
$\Lambda = 750$ GeV.  The $ZZ\gamma$ limits are similar.

\section{Prospects for the Near Future and for Run~II}
In the immediate future, D\O \ will combine the two new analyses with the 
previous, finalizing the Run~I anomalous $WW\gamma$ and $WWZ$ coupling limits.
Perhaps CDF will finish it's Run~I $W\gamma$ and $Z\gamma$ analyses soon.

Presently D\O \ and CDF are working on their detector upgrades for Run~II.
The Main Injector will allow the Tevatron to provide $2$ fb$^{-1}$ data samples
to each detector at $\sqrt{s}=2000$ GeV.  For D\O ,
the addition of a solenoid magnet and new tracking system will improve 
the muon resolution. The CDF detector will have improved electron and 
photon acceptance in the forward direction. These modifications strengthen
the detectors ability to study diboson final states.

Limits on anomalous couplings scale by 
approximately the 1/4 root of the luminosity for
fixed $\Lambda$ and assuming no improvement in technique. 
The large data samples will provide upwards of 3000 $W\gamma\rightarrow \ell 
\nu \gamma$ events, $700$ $Z\gamma \rightarrow ee\gamma + \mu\mu\gamma $ 
events, $100$ $WW\rightarrow {\rm dileptons}$ events, some 30 
$WZ\rightarrow {\rm trileptons}$ and a handful of $ZZ\rightarrow e's \; 
{\rm and } \; \mu's$ per experiment. 

\begin{figure}
\psfig{figure=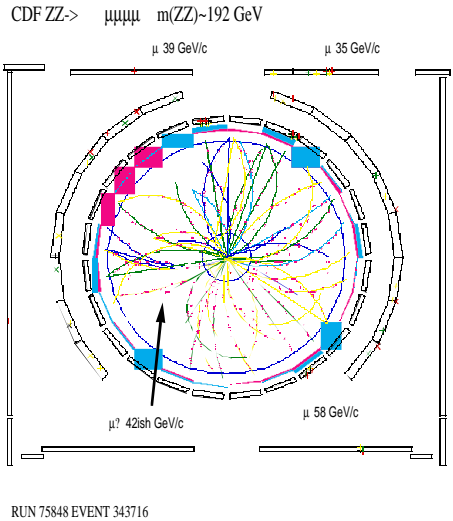,height=2.9in}
\caption{$ZZ\rightarrow \mu\mu$ event observed at CDF in Run~I. The $E_T$ and 
locations of the muons are shown. } 
\label{fig:cdfzz}
\end{figure}

CDF has already observed a $ZZ$ event during Run~I.
Figure~\ref{fig:cdfzz} shows their $ZZ\rightarrow \mu^+\mu^-\mu^+\mu^-$ event.
This is the first Z boson pair candidate recorded. They expected slightly
less than one such event in Run~I. 

Qualitatively, the $W\gamma$, and perhaps, the $WZ$ radiation zeroes will be 
unambiguously observed. Anomalous trilinear coupling limits will begin to 
probe the theoretical expectations. The first measurements  of quadrilinear 
couplings  will be available. 

\section{Summary}
Direct measurements of vector boson pair production processes and 
trilinear gauge boson couplings have been conducted by the CDF and D\O \ 
Collaborations.  Preliminary results from searches for 
anomalous $WW/WZ\rightarrow \mu\nu$ jet jet and 
$WZ\rightarrow {\rm eee}\nu$ production from D\O \ were presented. 
95\% CL anomalous coupling limits 
from previously published D\O \ results are $-0.20 < \lambda < 0.20 \; 
(\Delta\kappa=0)$ and $-0.30 < \Delta\kappa < 0.43 \; (\lambda=0)$ for 
$\Lambda=2000$ GeV where the $WW\gamma$ couplings are assumed to equal the 
$WWZ$ couplings.  
Combined D\O \ + LEP experiment anomalous coupling limits were presented for
the first time. 95\% CL limits are $-0.16<\lambda_{\gamma}<0.10\; 
(\Delta\kappa=0)$ and $-0.15<\Delta\kappa_{\gamma}<0.41\; (\lambda=0)$
under the assumption that the couplings are related  by the ``HISZ'' 
constraints. 
95\% CL anomalous $ZZ\gamma$ and $Z\gamma\gamma$ coupling limits from D\O \
are $|h_{30}^{Z}|<0.36 \; (h_{40}^{Z}=0)$ and $|h_{40}^{Z}|<0.05 \;
(h_{30}^{Z}=0)$ for $\Lambda=750$ GeV. 
CDF reports the first observation of a $ZZ$ event. 

Run~II, with upgrades to the CDF and D\O \ detectors and larger integrated 
luminosities due primarily to the Main Injector, will provide even more 
tantalizing opportunities for studying gauge boson couplings.  


\end{document}